\documentclass[fleqn,usenatbib]{mnras}

\usepackage{newtxtext,newtxmath}

\usepackage[T1]{fontenc}
\usepackage{ae,aecompl}
\usepackage[table]{}
\usepackage{textcomp}

\usepackage{graphicx}	
\usepackage{amsmath}	
\usepackage{hyperref}
\usepackage{tablefootnote}
\usepackage{wasysym}
\usepackage{float}
\usepackage{tabularx}
\usepackage{comment}
\usepackage{multirow}

\title[Framework for detecting small Solar System bodies in exoplanet surveys]{GPU-based framework for detecting small Solar System bodies in targeted exoplanet surveys}

\author[Burdanov et al.]{A.~Y.~Burdanov$^{1}$\thanks{E-mail: burdanov@mit.edu (AYB);  shasler@mit.edu (SNH); jdewit@mit.edu (JdW)},
S.~N.~Hasler$^{1}$\color{blue}\footnotemark[1]\color{black},
J.~de~Wit$^{1}$\color{blue}\footnotemark[1]\color{black}
\\
$^{1}$Department of Earth, Atmospheric and Planetary Sciences, Massachusetts Institute of Technology, 77 Massachusetts Avenue, Cambridge, MA 02139, USA\\}

\date{}

\pubyear{2022}

\begin{document}
\label{firstpage}
\pagerange{\pageref{firstpage}--\pageref{lastpage}}
\maketitle

\begin{abstract}

Small Solar System bodies are pristine remnants of Solar System formation, which provide valuable insights for planetary science and astronomy. Their discovery and cataloging also have strong practical implications to life on Earth as the nearest asteroids could pose a serious impact threat. Concurrently with dedicated observational projects, searches for small bodies have been performed on numerous archival data sets from different facilities. Here, we present a framework to increase the scientific return of an exoplanet transit-search survey by recovering serendipitous detections of small bodies 
in its daily and archival data using a GPU-based synthetic tracking algorithm. As a proof of concept, we analysed $12\,\times12\,\mathrm{arcmin^2}$ sky fields observed by the 1-m telescopes of the SPECULOOS survey. We analysed 90 sky fields distributed uniformly across the sky as part of the daily search for small bodies and 21 archival fields located within 5\,deg from the ecliptic plane as part of the archival search (4.4\,deg$^2$ in total). Overall, we identified 400 known objects of different dynamical classes from Inner Main-belt Asteroids to Jupiter Trojans and 43 potentially new small bodies with no priors on their motion. We were able to reach limiting magnitude for unknown objects of $V$=23.8\,mag, and a retrieval rate of $\sim$80\% for objects with $V<$22\,mag and $V<$23.5\,mag for the daily and archival searches, respectively. SPECULOOS and similar exoplanet surveys can thus serve as pencil-beam surveys for small bodies and probe parameter space beyond $V$=22\,mag.

\end{abstract}




\section{Introduction}\label{sec:Introduction}

Small Solar System bodies are rocky, icy and metallic objects orbiting the Sun, which are not massive enough to satisfy the definition of a planet or a dwarf planet\footnote{See International Astronomical Union (IAU) resolution B5 on "Definition of a Planet in the Solar System"}. This group of objects is very diverse (in terms of size, orbit and composition) and includes asteroids, comets, most Trans-Neptunian Objects (TNOs), and interplanetary dust particles. 

The study of small bodies is a broad topic and it holds a significant value in several areas of scientific discovery. Asteroids are planetesimal remnants related to the formation of the terrestrial and giant planets of the Solar System, while comets are remnants of the initial population of planetesimals of the outer part of the protoplanetary disk. Compared to larger Solar System bodies, asteroids and comets have been minimally affected since their formation and their orbits have been shaped by early dynamical processes. This means they contain an invaluable record of the early Solar System and its evolution \citep{2014Natur.505..629D, michel2015asteroids,2015SSRv..197..191D}. The presence of volatiles on some small bodies suggests that comets and asteroids could lead us to answers about the distribution of water in the Solar System and how Earth gathered an abundance of water to support life \citep{osinski2020role}. Finally, some small bodies could also pose an impact threat to Earth \citep{michel2015asteroids}, and their discovery and cataloging have strong practical implications \citep{board2010defending}. 

Major developments in our understanding of small Solar System bodies have been made thanks to all-sky wide-field observational projects \footnote{\url{https://minorplanetcenter.net/iau/lists/MPDiscsNum.html} (Accessed 2023 February 21)}, such as Spacewatch \citep{1982ASSL...96..279G}, Lincoln Near-Earth Asteroid Research (LINEAR; \citealt{stokes2000lincoln}), Catalina Sky Survey
(CSS; \citealt{larson1998catalina}),  Near-Earth Asteroid Tracking (NEAT; \citealt{1999AJ....117.1616P}), and Panoramic Survey Telescope and Rapid Response System (Pan-STARRS; \citealt{kaiser2002pan}), as well as space-based surveys such as the Wide-field Infrared Survey Explorer mission (WISE; \citealt{wright2010wide}). Simultaneously, searches for specific populations of small bodies are accomplished by dedicated surveys, e.g., TNO surveys \citep{bannister2016outer, weryk2016distant, chen2018searching, sheppard2019probing} and surveys for bodies located inside the orbit of Venus \citep{2022MNRAS.517L..49B}.

Currently, the IAU Minor Planet Center (MPC) lists $\sim$1.3 million discovered small bodies\footnote{\url{https://minorplanetcenter.net} (Accessed 2022 October 01)}, where the overwhelming majority is the population of main-belt asteroids (MBAs). Many more discoveries of small bodies of different classes will be driven in the future by next generation projects such as the Rubin Observatory \citep{2009arXiv0912.0201L}, Roman Space Telescope \citep{2018JATIS...4c4003H}, NEOCam \citep{2017DPS....4921901M} and Euclid \citep{2018A&A...609A.113C}. 

Concurrent with dedicated observational projects, searches for small bodies have been performed on various archival data sets \citep{vaduvescu2009euronear, 2012PASP..124..579G,2020A&C....3000356V}. Indeed, many astronomical observations contain serendipitous detections of small bodies. Mining such data can be a useful tool for making statistical inferences about their population \citep{2001AJ....122.2749I,2016A&A...591A.115P,2021A&A...652A..59S}, gaining various insights from small body photometric light curves \citep{2020Icar..34513721M} and discovering new objects  \citep{2022ApJS..258...41B}. In this regard, space-based missions hold a special place as they offer all-sky coverage and/or increased sensitivity to fainter objects. For example, \cite{2022A&A...661A..85K} used \textit{Hubble Space Telescope (HST)} data to detect numerous asteroids as faint as $V\sim$24.5\,mag. \cite{2021PASP..133a4503W} developed a pipeline for the \textit{Transiting Exoplanet Survey Satellite} (\textit{TESS}; \citealt{2015JATIS...1a4003R}) to detect comets, asteroids and near-Earth objects (NEOs) as faint as $V\sim$19.5\,mag using conventional techniques. A rich set of small bodies down to $G\sim$21.0\,mag from \textit{Gaia} astrometric mission was reported by \cite{2022arXiv220605561T}. 

Small bodies are identified in images due to their motion on the sky relative to the background stars. This motion is a combination of the actual orbital movement of the body and Earth’s reflex motion. Most of the above-mentioned projects rely on detecting objects in single images and linking detected motion between multiple exposures. This method is robust and computationally fast, but it has its limitations for fainter objects (i.e., smaller and/or more distant), which might not show up in a single exposure. Detecting such objects would require a more sensitive telescope/detector, but it can be mitigated in certain cases by combining multiple existing exposures. 

The "shift-and-stack" technique, first introduced in the 1990s \citep{tyson1992limits, cochran1995discovery}, can improve the signal-to-noise ratio (S/N) of small bodies by combining multiple exposures. This technique is accomplished by removing the high-S/N stationary sources from each image, predicting the linear motion of the object of interest, shifting the image pixels with respect to the motion, and then combining the shifted images. This results in an image where the flux from only the moving object has been added constructively \citep{parker2010pencil}. However, the original "shift-and-stack" technique requires prior knowledge of the object's motion on the sky. Synthetic (or digital) tracking is an extension of the "shift-and-stack" method that does not require prior knowledge of an object's motion and can be employed as a real-time tracking and identification technique \citep{shao2014finding}. The technique has been proven to be successful in several dedicated ground-based small body surveys \citep{2015AJ....150..125H,2020PASP..132f4502Z} and in the \textit{TESS} exoplanet survey, enabling the recovery of objects as faint as $V\sim$21.0\,mag with no prior information about objects' motion \citep{2020PSJ.....1...81R}. Its main drawback of computational cost can be alleviated by performing computations using graphics processing units (GPUs).

In this paper, we present the first results of a framework to increase the scientific potential of an exoplanet transit survey by recovering serendipitous detections of small bodies in its photometric data. We developed a pipeline based on the publicly-available GPU-accelerated synthetic tracking software \textsc{Tycho Tracker} \citep{2020JAVSO..48..262P}. We extensively tested the pipeline on archival and last-night data sets from the SPECULOOS (Search for Planets EClipsing ULtra-cOOl Stars) targeted ground-based photometric survey, which aims to find terrestrial exoplanets transiting the nearest ultra-cool dwarf stars. 

The rest of the paper is organized as follows. In Section~\ref{sec:pipeline}, we present the data processing and analysis pipeline. Section~\ref{sec:data} describes information about the SPECULOOS survey and the data it provides. We review application of the pipeline to last-night and archival images from SPECULOOS in Section~\ref{sec:discuss}, and we summarise our findings in Section~\ref{sec:conclusions}.

\begin{figure*}
    \centering
    \includegraphics[width=\textwidth]{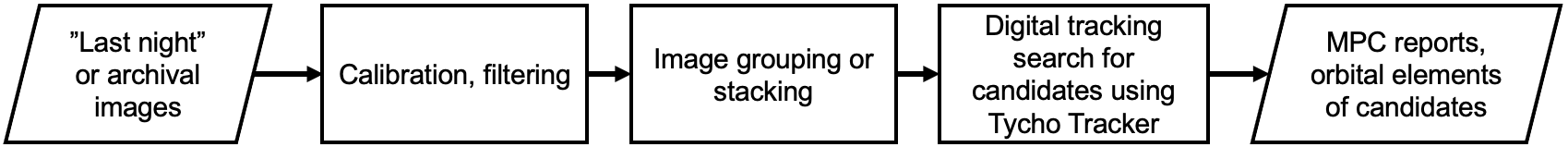}
    \caption{Data processing and analysis pipeline structure. Both the archival and last-night data sets are processed following the same structure. Raw FITS images are first calibrated and filtered. Then, images are grouped or stacked based on the type of search being performed and then passed through to the synthetic tracker. RA and DEC measurements of every detection (matched and unmatched to known objects) are generated for submission to MPC. Orbital parameters of objects of interest are derived for follow-up observations.}
    \label{fig:pipeline_flowchart}
\end{figure*}

\section{Data processing and analysis pipeline}\label{sec:pipeline}

We developed two versions of the data processing and analysis pipeline. One version is designed to run daily after the end of an observing night and provides a list of all possible moving object candidates before the beginning of the next observing night. We refer to it as the last-night data pipeline. Its development is motivated by our desire to submit new observations to the MPC in a timely manner and to be able to follow up objects of interest during the next observing night. The other version of the pipeline is intended to process archival images and is focused on detecting small bodies that move relatively slowly and could be present within the telescope's Field of View (FoV) for several subsequent nights (which greatly improves orbit determination). We refer to this version as the archival data pipeline. 

Both versions of the pipeline are run on a computer with a \textsc{Windows~10} operating system. All parts of the pipeline are coded in \textsc{Python} programming language, except the synthetic tracking algorithm. It is realised in \textsc{Tycho Tracker\footnote{\url{https://www.tycho-tracker.com} (Accessed 2022 February 21)}}, which communicates with GPUs using the \textsc{OpenCL} framework. We present more a detailed description of the pipelines in the following subsections and outline the data pipeline structure in Fig.~\ref{fig:pipeline_flowchart}. 

\subsection{Last-night data}

The last-night data are raw FITS files that must be calibrated and corrected accordingly before they can be searched for moving objects. We perform calibration following a standard technique (bias, dark and flat-field correction) for all last-night data sets using \textsc{ccdproc}, an \textsc{Astropy} affiliated \textsc{Python} package \citep{2013A&A...558A..33A, 2018AJ....156..123A}. Then, we perform additional correction steps, including masking and removing bad pixels. We remove cosmic rays using the  \textsc{LACosmic} method \citep{2001PASP..113.1420V, curtis_mccully_2018_1482019} and align the images relative to one another using the \textsc{Astroalign Python} module \citep{beroiz2020astroalign}. Finally, astrometric calibration of images is done using \textsc{astrometry.net} \citep{2010AJ....139.1782L}.

In the next step, the pipeline performs source extraction using the image segmentation technique realised in \textsc{Photutils}, an \textsc{Astropy} affiliated \textsc{Python} package. Different parameters of images and extracted sources are analysed in order to filter the images. Defocused and satellite-contaminated images are discarded based on mean FWHM, where FWHM is the mean full width at half maximum of the stars' point spread function (PSF) in the image. Images obtained in poor weather conditions (e.g., clouds) are filtered based on the number of detected stars and on the standard deviation of the sky pixel counts.

Calibrated and filtered images from each telescope pointing (sky field) are passed to \textsc{Tycho Tracker} in several groups. The grouping of images is motivated by the fact that a moving object can be detected by the software only if it is present above a given threshold for at least half of the images. This detection limitation does not pose a problem for slowly moving objects, which are present in the FoV throughout the whole image sequence. However, faster moving objects in long image sequences might be lost (e.g., if 100 images are passed to \textsc{Tycho Tracker} and an object is visible only on 20 images). To minimise the possible loss of fastest detectable objects, an initial image sequence is divided into a number of slightly overlapping groups which are passed to synthetic tracker one by one. Here, speed of the fastest detectable objects is defined by the exposure time --- a small amount of streaking per exposure due to object's movement is allowed (several of typical FWHM of the images). Image group size is defined in such a way that the fastest detectable objects is present in the group for 70\% of the time. After the search for fast moving objects is completed, the pipeline sums every $N$ images from the observing run so that the resulting summed images have a total exposure of several minutes. Such summed image sequence is passed to \textsc{Tycho Tracker} to search for slow-moving objects.

\begin{figure}
    \centering
    \includegraphics[width=\columnwidth]{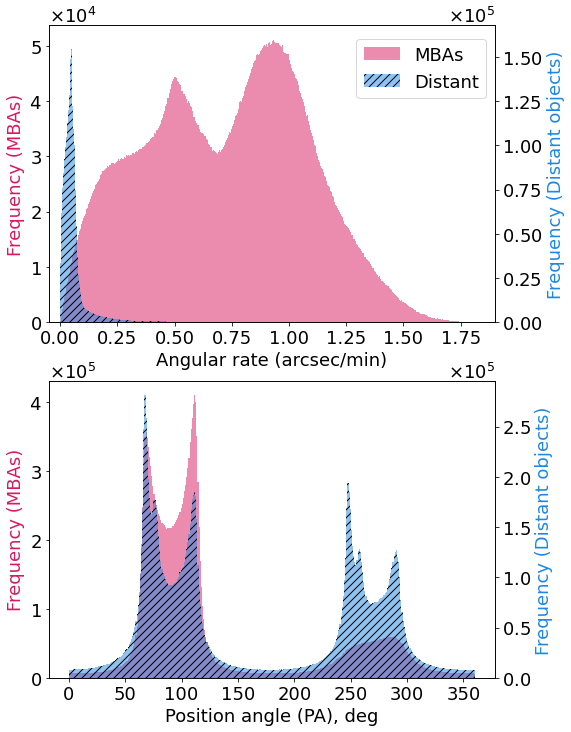}
    \caption{Geocentric total apparent angular rate in the plane-of-sky and position angle (PA) distributions for a random 1\% of Main-belt Asteroids (MBAs) and for all Distant objects (TNOs, Centaurs and Scattered Disk Objects) from the MPC database. The distributions were created by querying JPL Horizons ephemeris system (accessed 2022 August 1; \citealt{1996DPS....28.2504G}) for every object with a start date of 1 Apr 2017, end date of 1 Apr 2027, and a time step of 1\,day. The PA of the target's direction of motion in the plane-of-sky is measured counter-clockwise from the apparent of-date north pole direction. PA peaks around 90 and 270\,deg reflect motion of objects before and after opposition, respectively.}
    \label{fig:speed_PA_distrib}
\end{figure}

For every group of images, the median combination of the image sequence is used to remove the high-S/N stationary sources (stars) from each image. Though small bodies have a set of predominant position angles (PAs) and angular speeds (see Fig.~\ref{fig:speed_PA_distrib}), we perform shifting and summing on cleaned images with no constraints on PA under the assumption of their linear motion.

As mentioned above, the maximum speed limit is defined by the exposure time. The minimum speed limit is defined by the minimum shift of the images (usually, 0.5 pixel). After shifting and stacking, \textsc{Tycho Tracker} analyses images where the fluxes from only the moving objects have been added constructively. A set of candidate detections (tracks) is returned with corresponding speed, PA, pixel coordinates and S/N of detection.

When detections are made, the MPC database of small body orbits and \textsc{Find\_Orb}\footnote{\url{https://www.projectpluto.com/find_orb.htm} (Accessed 2022 November 01)} software are used by \textsc{Tycho Tracker} to cross-match the detections with already known objects. A match is made if a known object is within a certain distance of the candidate detection. For our pipeline, we restricted a match distance (observed minus calculated coordinates) to 0.3\,arcmin to minimise false matches. RA and DEC measurements of every matched and unmatched candidate (``tracklets'') are generated by \textsc{Tycho Tracker} and can be further sent to the MPC. We also use the tracklets to fit the orbit using \textsc{Find\_Orb} and generate ephemeris for follow-up observations. We present specifics of archival image processing in the next subsection.  

\subsection{Archival data}

Generally, if a small body was observed during only a short, one-night arc and was not followed up afterwards, its orbit determination will be poor. Such objects might be lost completely or until future recovery observations. Objects which happen to be in an FoV for several consecutive nights might offer better prospects of orbit determination (typically, tracklets from three different nights are needed for robust MBA orbit determination; see \citealt{2018arXiv180502638H} and references therein). Such objects will be also better placed for detection by synthetic tracking as more images will be stacked together, increasing the objects' S/N (and enabling probes of more distant populations of small Solar System bodies). Motivated by this, we focused on detecting slow-moving objects, treating archival sky fields as "deep drilling" data sets.

For every sky field, the archival pipeline performs data calibration and filtering similar to the last-night version. Then, instead of image grouping, it sums every $N$ images for every night so that the resulting summed images have a total exposure of several minutes. Broadly, $N$ is defined so that a slow-moving object moves more than 0.5 pixels from one exposure to another. $N$ depends on the exposure time of individual images, the image scale, and the upper speed limit of the object (typically, total apparent angular rate in the plane-of-sky $<$1\,arcsec/min). For a typical observation with 30\,s exposures, $N$=14 what results in a summed image with 420\,s exposure time.

\begin{figure}
    \centering
    \includegraphics[width=\columnwidth]{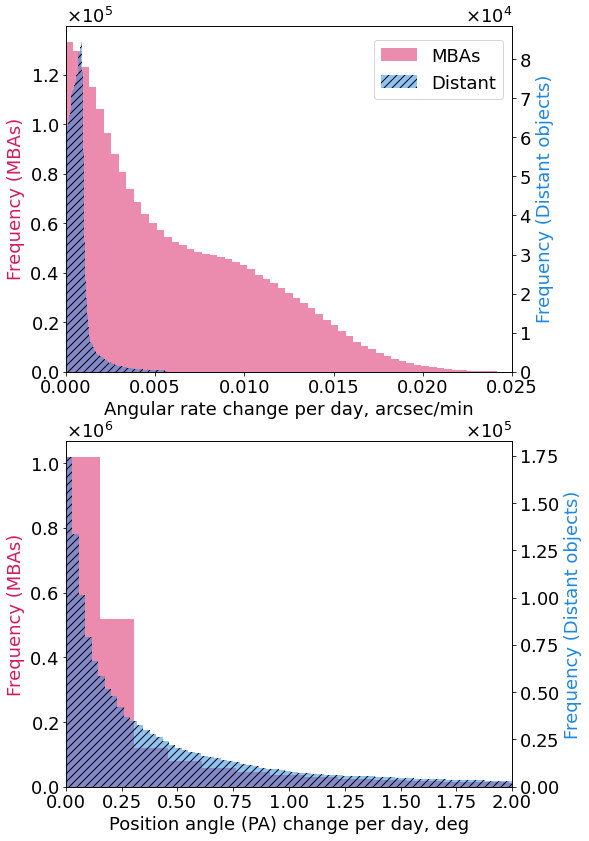}
    \caption{Daily changes of geocentric total apparent angular rate in the plane-of-sky and position angle (PA) distributions for a random 1\% of Main-belt Asteroids (MBAs) and for all Distant objects (TNOs, Centaurs and Scattered Disk Objects) from the MPC database. The distributions were created by querying JPL Horizons ephemeris system (accessed 2022 August 1) for every object with a start date of 1 Apr 2017, end date of 1 Apr 2027 and a time step of 1\,day. Presented PA change rates are limited to 2 deg per day.}
    \label{fig:speed_change}
\end{figure}

In the next steps, a set of summed images is passed to \textsc{Tycho Tracker} for synthetic tracking. Detected candidates are linked over several nights and their orbital parameters are derived with the \textsc{Find\_Orb} software. We note that after each individual night from a particular sky field is processed, the archival data pipeline also performs synthetic tracking on a group of images comprising several consecutive nights together (referred to as 2-night data set hereafter). In this case, objects showing non-linear motion due to significant speed and PA changes from one night to another will not be detected, as the synthetic tracker cannot use a single motion vector for them. However, depending on their orbital configuration, a large fraction of small bodies may show such a linear motion and could be successfully detected using synthetic tracking. To illustrate this statement, we computed daily changes of PA and speed of motion in the plane-of-sky for all known MBAs and distant objects presented by the MPC (TNOs, Centaurs and Scattered Disk Objects) over the course of 10 years. As shown in Fig.~\ref{fig:speed_change}, a significant fraction of MBAs and distant objects have daily changes of PA smaller than 0.5\,deg and daily changes of speed smaller than 0.005\,arcsec/min. This allows successful recovery of such objects with synthetic tracking over the course of several nights, which will be demonstrated in Section~\ref{sec:discuss}. 

\subsection{Injection-retrieval tests}\label{subsec:inj-retr}

The most time-consuming part of the data processing pipeline is synthetic tracking. Here, the processing time scales quadratically with the number of images passed to the synthetic tracker and linearly with the size of the images (number of pixels). In order to reduce processing time to have candidate detections before the beginning of the next observing night and not to lose dramatically in the sensitivity to faint and fast objects, we experimented with different image grouping, spatial down-sampling (binning) and image summing. To this end, we ran series of injection-retrieval tests. We injected 1000 synthetic moving objects in raw FITS files representing a typical observing night from a facility of interest. Objects' FWHM in each image equals mean FWHM of that image derived from the analysis of stars. Our synthetic moving object injection module uses real distributions of small bodies angular speed, PA (Fig.~\ref{fig:speed_PA_distrib}) and visual apparent magnitudes (Fig.~\ref{fig:ap_mag_distr}). After processing the data with the pipeline, detected objects were compared with the injected ones (see also an example of the results for injection-retrieval by object magnitude in Fig.~\ref{fig:inj-retr_mag}). 

For the typical observing night of 5 hours with 25\,s exposure time from the SPECULOOS telescopes (see Section~\ref{sec:data}), we found these parameters to be the optimal: group size of 50 images and $2\times2$ binning for detection of fast-moving objects, and summing of every 13 images and no binning for detection of slow-moving objects. We provide a detailed description of the archival and last-night data sets from the SPECULOOS telescopes in the next section.

\begin{figure*}
    \centering
    \includegraphics[width=0.9\textwidth]{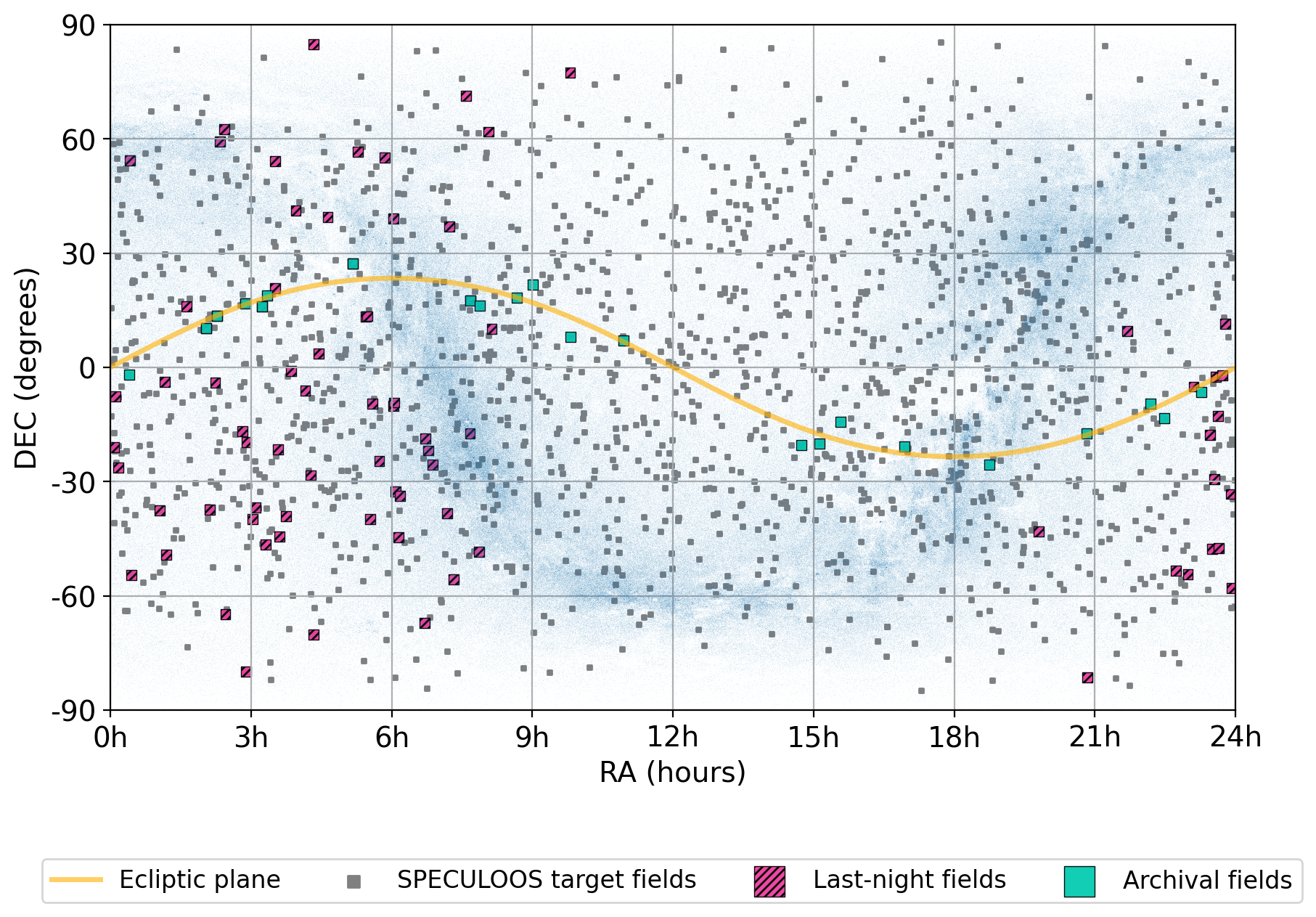}
    \caption{Projection of a celestial sphere with the ecliptic plane (orange line) over-plotted and stars (blue dots) from the Tycho-2 catalogue \citep{2000A&A...355L..27H}. Searched SPECULOOS archival fields (teal squares) and last-night fields (pink hatched squares) are also shown. The size of all the fields on the plot is exaggerated for better viewing.} 
    \label{fig:FOVs_distribution}
\end{figure*}

\section{Data}\label{sec:data}

SPECULOOS is a ground-based targeted transit survey which aims to detect terrestrial exoplanets orbiting the nearest ultra-cool dwarfs \citep{2018NatAs...2..344G,2018Msngr.174....2J,2018SPIE10700E..1ID}. It is composed of six identical 1-m Ritchey-Chr\'etien robotic telescopes, equipped with $\mathrm{2K\times2K}$ CCD detectors with a pixel size of 13.5~$\mu$m. The FoV of each telescope is $12\,\times12\,\mathrm{arcmin^2}$ and the corresponding pixel scale is 0.35\,$\mathrm{arcsec\,pixel}^{-1}$. Since 2019, all telescopes have been photometrically monitoring targets from the SPECULOOS input catalog, aiming to collect between 100 and 200 hours for every target in the near-infrared. There are 1700 targets in the input catalog, which are distributed uniformly across the sky \citep{,2021A&A...645A.100S}. Upon completion of the survey, all the telescopes will have observed almost 70\,$\mathrm{deg^2}$ of sky with remarkable photometric precision and high cadence \citep{2020MNRAS.495.2446M, burdanov2022speculoos}. 

Each night, every SPECULOOS telescope typically observes 1--2 targets from the input catalog, mostly in the $I+z'$ filter \citep{2018SPIE10700E..1ID} with exposure times ranging from 10 to 120\,s. Occasionally, various follow-up targets are also observed (e.g., transiting exoplanet candidates from the \textit{K2} and \textit{TESS} missions; \citealt{Niraula2020, Wells2021, 2022A&A...667A..59D,2022A&A...657A..45S}). These images are then processed daily to extract the photometry of all static point sources in an FoV using the independent photometric pipelines described in \citealt{2020MNRAS.495.2446M}, \citealt{2020A&A...642A..49D} and \citealt{2022MNRAS.509.4817G}. Our last-night data pipeline for moving targets in an FoV runs daily on a dedicated computer with two NVIDIA RTX\texttrademark A6000 GPUs and is optimised to process data before the start of the next night for each telescope. For the sake of the testing, the pipeline has been processing data only from four telescopes from mid-September to mid-November 2022: SPECULOOS-Artemis, installed at the Teide Observatory (Spain; MPC observatory code Z25); and SPECULOOS-Europa, SPECULOOS-Io and SPECULOOS-Ganymede, installed at the ESO Paranal Observatory (Chile; MPC observatory code W75). 

\begin{table}
    \centering
    \begin{tabular}{lc}
    \hline
        Orbital grouping & Orbital parameters \\
    \hline
        Near-Earth Asteroid & $q < 1.3$ \\
        Inner Main-belt Asteroid & $a < 2.0, q > 1.666$ \\
        Main-belt Asteroid & $2.0 < a < 3.2, q > 1.666$ \\
        Outer Main-belt Asteroid & $3.2 < a < 4.6$ \\
        Mars-crossing & $a < 3.2, 1.3 < q < 1.666$ \\
        Jupiter Trojan & $4.6 < a < 5.5, e < 0.3$ \\
        Comet-like & $Q > 5.0$ \\
    \end{tabular}
    \caption{Small bodies are assigned to groups and families based on their orbital elements, where $a$ is the semi-major axis (AU), $e$ is the orbital eccentricity, $i$ is the orbital inclination (°), $q$ is the perihelion distance (AU), and $Q$ is the aphelion distance (AU).}
    \label{tab:orb_params}
\end{table}

Depending on the exposure time, typical datasets from each night comprises several thousand FITS images. Defocused images  ($\mathrm{FWHM}>10$ pixels) and those observed in cloudy conditions (featuring $<10$ stars) are discarded. Detailed information about fields processed by the last-night version of the pipeline with small body detections (26 fields) is presented in Table~\ref{slow_search_fields} and all 90 observed fields' sky-projected positions are presented in Fig.~\ref{fig:FOVs_distribution}.

At the time of writing, the SPECULOOS telescopes have accumulated 300 targets from the input catalog with more than 10 hours each. To test our archival data pipeline, we selected 21 sky fields which are located within 5\,deg from the ecliptic plane (to increase our chances of detecting small bodies) and which were observed for more than 10 hours (and at least 7-8 hours per night), starting from the commissioning of the telescopes in 2018 until mid-2022 (see Table~\ref{slow_search_fields} and Fig.~\ref{fig:FOVs_distribution}). As in the case for the last-night data, the archival pipeline processing uses GPUs (four NVIDIA RTX\texttrademark 3090). In the next section, we review and discuss detections from each version of the pipeline.

\section{Results and Discussion}\label{sec:discuss}

\begin{figure}
    \centering
    \includegraphics[width=\columnwidth]{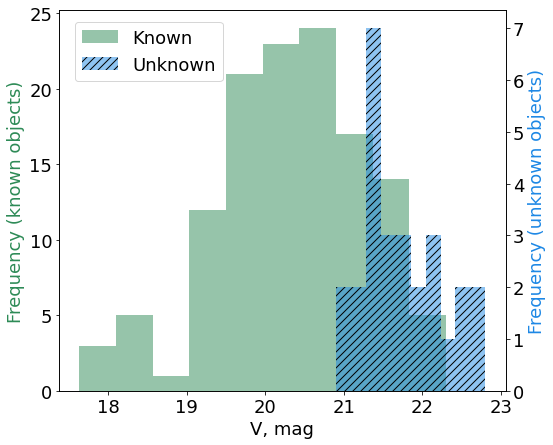}
    \caption{$V$\,mag distribution of known and unknown objects detected by the last-night data pipeline.}
    \label{fig:last_night_Vmag_known_objects}
\end{figure}

\begin{figure}
    \centering
    \includegraphics[width=\columnwidth]{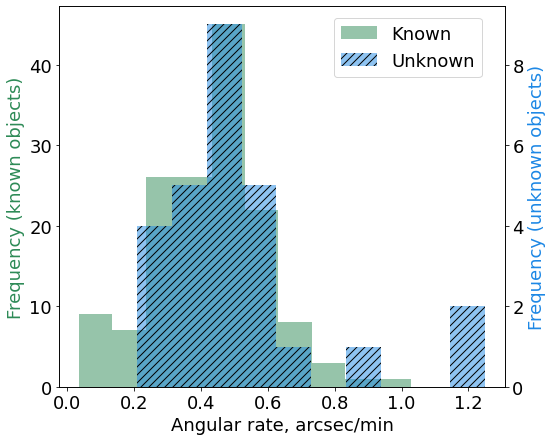}
    \caption{Total apparent angular rate in the plane-of-sky of known and unknown objects detected by the last-night data pipeline.}
    \label{fig:last_night_known_speed_pa_distr}
\end{figure}

First, we review and discuss results from the last-night data pipeline. Over the course of 60 nights, 90 different fields (3.6\,deg$^2$ in total) were observed with four SPECULOOS telescopes, mostly in $I+z'$ filter and occasionally in $g', r', i'$ and $z'$ filters. We detected 148 known objects in 24 fields (0.96\,deg$^2$) with a median matching distance of 1.5\,arcsec. These objects belong to different dynamical classes, including: three Inner Main-belt Asteroids (IMBs), two Mars-crossing asteroids, 134 Main-belt Asteroids (MBAs), four Outer Main-belt Asteroids (OMBs), four Jupiter Trojans (TJNs) and one comet. Dynamical classes of detected known objects were retrieved from NASA/JPL Small Body Database\footnote{\url{https://ssd-api.jpl.nasa.gov/doc/sbdb.html} (Accessed 2023 February 21)} (see Table~\ref{tab:orb_params}). The distribution of $V$\,mag of these objects is presented in Fig.~\ref{fig:last_night_Vmag_known_objects} and their total apparent angular rate in the plane-of-sky is presented in Fig.~\ref{fig:last_night_known_speed_pa_distr}. The brightest detected object has $V$=17.6\,mag, the faintest $V$=22.3\,mag, and the median value of $V$\,mag distribution is 20.4\,mag.

Additionally, 100 detections with S/N\,$>$\,15 in shifted and stacked images could not be matched to any known object closer than 0.3\,arcmin by \textsc{Tycho Tracker}. We also checked positions of any known objects using \textsc{MPChecker} tool and could not find any matches. Taking into account detections of the same object in different groups of images, 27 unique objects were identified, ranging in brightness from $V$=20.9\,mag to $V$=22.8\,mag. The median value of $V$\,mag distribution of unknown objects is 21.6\,mag. Their speeds ranged from 0.2 to 1.25\,arcsec/min, very likely corresponding to MBA orbits. These detections were made in 14 fields (in 12 of them, we also detected known objects). Overall, we detected 175 small bodies (148 known and 27 unknown) in 26 fields (1.04\,deg$^2$) located within 30\,deg from the ecliptic plane by the last-night version of the pipeline.

To assess efficiency of the pipeline, we compared detected known small bodies with those predicted to be within an FoV at the time of observations. We used the NASA Jet Propulsion Laboratory (JPL) Small-Body Identification\footnote{\url{https://ssd-api.jpl.nasa.gov/doc/sb_ident.html} (Accessed 2023 February 21)} Application Program Interface (API) to obtain small bodies' orbits (which were numerically integrated using a high-fidelity force model). According to the ephemeris, 217 small bodies were present in the last-night FoVs with speeds in the 0.01-2.3\,arcsec/min range and brightness ranging from $V$=17.5\,mag to $V$=30.0\,mag (with a median value $V$=21.4\,mag). Our pipeline was able to detect 81\% of all possible known objects down to $V$=22.3\,mag (faintest detected known object). Three Near-Earth Asteroids (NEAs) were present in FoVs, but they were not detected because of their faintness ($V\geq23.7$\,mag).

We note that the fastest detected object has a speed of 1.25\,arcsec/min. Generally, objects as fast as 8\,arcsec/min can be detected by the SPECULOOS telescopes, given a minimum exposure time of 10\,s. This range covers typical speeds of NEAs. \citet{2018AJ....156....5V} found that delayed submission of newly observed NEA candidates to the MPC plays a significant role in a large fraction of the unconfirmed NEA candidates (which might have MBA-like motions). A relatively quick processing time and ability to submit MPC reports by the last-night pipeline could allow for timely follow-up of NEAs and other types of small bodies and minimise losses of such objects.

\begin{figure}
    \centering
    \includegraphics[width=\columnwidth]{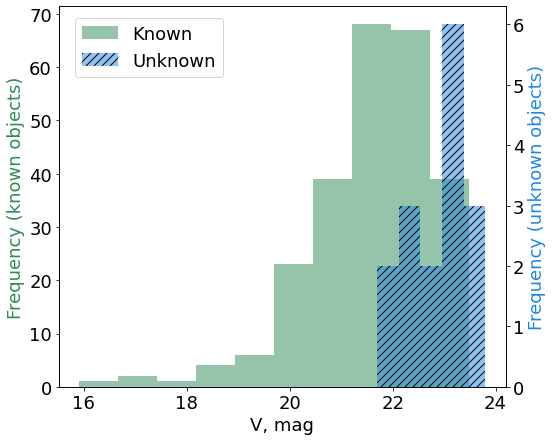}
    \caption{$V$\,mag distribution of known and unknown objects detected by the archival data pipeline.}
    \label{fig:archival_Vmag_knonwn_unknown}
\end{figure}

\begin{figure}
    \centering
    \includegraphics[width=\columnwidth]{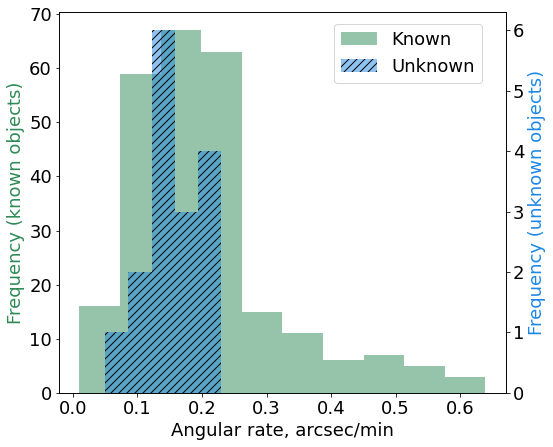}
    \caption{Total apparent angular rate in the plane-of-sky distribution for known and unknown objects detected by the archival data pipeline.}
    \label{fig:known_archival_speed_pa_distr}
\end{figure}

Regarding the search in the archival data, the pipeline processed 21 fields (0.84\,deg$^2$) --- that is, 390 individual nights and 240\,000 unsummed images --- in $\sim$7~days. It resulted in detections of 252 known objects in individual nights ranging in brightness from $V$=16.0\,mag (MBA Wesson) to $V$=23.5\,mag with a median matching distance of 0.7\,arcsec. Among them, 78 objects were detected on more than one night, with an extreme case of MBA 2002~QR57 ($V\sim$20\,mag). It was present in the FoV for 7 nights in November 2018 with a speed in 0.1-0.06\,arcsec/min range and PA in 270-10\,deg range. The vast majority of the known detected objects are MBAs (243 objects), with some detections of OMBs (8 objects) and one TJN. Semi-major axes $a$ of detected objects span a range from 2.1\,au to 5.2\,au. When the pipeline processed two consecutive nights as one set, 35 known objects were detected (MBAs: 32, OMBs: 2, TJN: 1). We note that these 35 targets were also detected in individual nights, however one object (MBA 1999~BP4) was identified in 2-night data sets only. The distribution of $V$\,mag of known detected objects is presented in Fig.~\ref{fig:archival_Vmag_knonwn_unknown} and their total apparent angular rate in the plane-of-sky in Fig.~\ref{fig:known_archival_speed_pa_distr}. The median value of $V$\,mag distribution of known objects detected by the archival pipeline is 21.8\,mag.

As in the case of the last-night data, we assessed efficiency of the archival pipeline by comparing detected known small bodies with those predicted to be within the FoVs. According to the ephemeris, 896 small bodies were present in the FoVs with speed between 0.006-1.5\,arcsec/min range and brightness ranging from $V$=15.7\,mag to $V$=34.0\,mag (with a median value $V$=21.8\,mag). Because of the employed image summing, the archival pipeline is more sensitive to slowly-moving objects, as fast-moving objects appear streaked on summed images (median value of the angular speed of all detected known objects is 0.19 arcsec/min). If we consider only small bodies present in the FoVs with a speed below 0.3\,arcsec/min, then our pipeline recovery rate is 80\% for objects down to $V$=23.5\,mag (faintest detected known object). This result is in agreement with the injection-retrieval tests where objects with speed of 0.2\,arcsec/min where injected (see~Fig.~\ref{fig:inj-retr_mag}). Most of the undetected known objects spent too small amounts of time with the FoVs to be detected and/or were close to stars or image artefacts which prevented their robust detection. Regarding the NEAs, 22 objects were in the FoVs during the observations. None of them were detected as 20 of the objects were too faint ($V\geq24.1$\,mag) and the remaining 2 brighter objects were too fast (speed $\sim1$\,arcsec/min).

Additionally, 26 detections with S/N\,$>$\,15 in the archival data could not be matched to any known object by \textsc{Tycho Tracker} nor by the \textsc{MPChecker} tool. The brightest among them has $V$=21.7\,mag, and the faintest $V$=23.8\,mag (detected on 215 shifted and stacked images with 75\,s exposure time each). The median $V$\,mag value of these detections is 23.0\,mag (1.2\,mag fainter than for the known objects). Some of the objects were visible on consecutive nights and their positions were linked, leaving us with 16 new and unique identified targets. Three unknown objects were visible on three consecutive nights which allowed us to determine their orbital elements and attribute them to MBAs with semi-major axis $a$ in 2.3-3\,au range and inclination $i$ in 1-8\,deg range. 

\begin{figure}
    \centering
    \includegraphics[width=0.85\columnwidth]{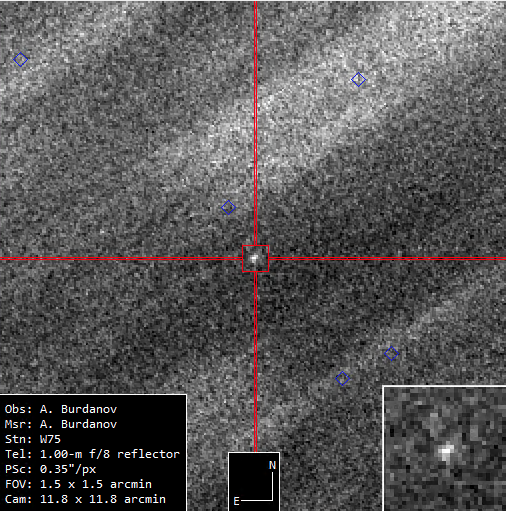}
    \caption{Detection of the faintest ($V$=23.8\,mag) unknown object in the archival data. Stacked and shifted image is composed of 215 images in $I+z'$ filter with 72\,s exposure time and obtained by the SPECULOOS-Ganymede telescope on 19 March 2020 in Sp1444-2019 field. The target's speed and PA are 0.23\,arcsec/min and 302\,deg respectively. The target's S/N on the shifted and stacked image is 15. Blue diamonds mark the locations of known stars from the Gaia~DR2 star catalog.}
    \label{fig:archival_faintest_unknown}
\end{figure}

\begin{figure}
    \centering
    \includegraphics[width=0.9\columnwidth]{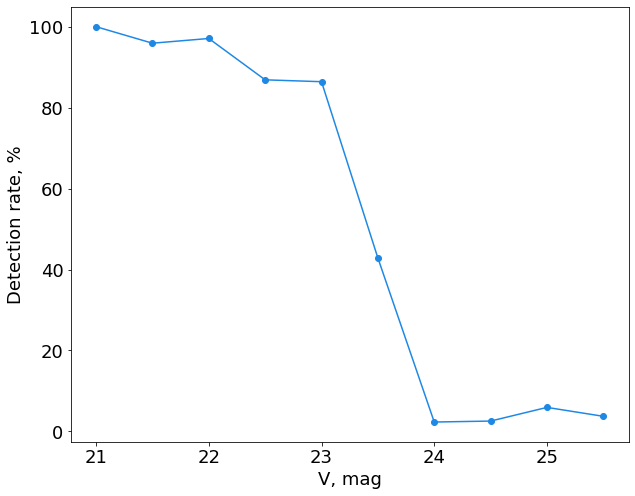}
    \caption{Injection-retrieval tests for \textbf{slow-moving targets (0.2\,arcsec/min)} performed on the data obtained by the SPECULOOS-Ganymede telescope on 19 March 2020 in Sp1444-2019 field where the faintest unknown target was detected.}
    \label{fig:inj-retr_mag}
\end{figure}

As mentioned above, the faintest and unknown detected object has $V$=23.8\,mag. Its stacked and shifted image is presented in Fig.~\ref{fig:archival_faintest_unknown}. It is composed of 215 images in $I+z'$ filter with 72\,s exposure time and obtained by the SPECULOOS-Ganymede telescope on 19 March 2020 in Sp1444-2019 field. The target's speed and PA are 0.23\,arcsec/min and 302\,deg respectively. To understand archival pipeline search completeness for this extreme case, we ran a series of injection-retrieval tests of slow-moving objects (0.2\,arcsec/min). We were able to recover 85\% of objects brighter than $V\approx23.0$\,mag, which is remarkable given that the data comes from 1-m optical telescope (Fig.~\ref{fig:inj-retr_mag}).

As expected, the archival pipeline tends to detect objects which move more slowly across the sky because of the employed image summing over a long observing night: median value of the angular speed of all known objects detected by the last-night pipeline is 0.45\,arcsec/min and 0.19\,arcsec/min for the known objects from the archival search. Similarly, archival search tends to detect fainter objects: the median $V$\,mag value of known object detections in the archival search is 21.8\,mag, which is 1.4\,mag fainter than the median value from the last-night search.

Several candidate detections of unknown objects in 2-night data sets were identified. However, they all were rejected after further inspection. We suspect that these candidate detections arose from coherent noise, which was removed when images with $2\times2$ binning were passed to \textsc{Tycho Tracker}. Nevertheless, this type of processing can be considered successful as MBA 1999~BP4 ($V\sim$21\,mag) was recovered only in 2-night sets because of its small apparent motion of only 0.01\,arcsec/min during its observations on 29 May 2022. This detection serves as a proof of concept for future application of our pipeline to detect TNOs and other distant objects with a typical speed of 0.01\,arcsec/min and a daily rate of PA change smaller than 0.25\,deg (see Fig.~\ref{fig:speed_PA_distrib} and Fig.~\ref{fig:speed_change}).

To put our results in context, we compared our findings from the last-night pipeline with the work of \citet{2019MNRAS.490.3046C}, who reported small body detections in four fields observed by the WFCAM Transit Survey (WTS). The survey operated on 3.8-m UKIRT telescope and targeted four fields of 1.6\,deg$^2$ each (three of the fields were located within 30\,deg from the ecliptic plane) with 5-10\,s exposures over 600 observing nights. In a total of 6.4\,deg$^2$, the authors detected 1821 different small bodies (285 objects per 1\,deg$^2$) using a conventional technique, i.e, detecting an object's displacement from image to image. Their reported limiting magnitude of known objects is $V$=21.5\,mag with the median value of $V\approx$19.5\,mag (see Fig. 3 from \citealt{2019MNRAS.490.3046C}). Though it is not trivial to compare different surveys given their difference in observing strategy and instrumentation, we detected a comparable number of small bodies in the data from SPECULOOS 1-m optical telescopes --- 175 objects in 1.04\,deg$^2$. In contrast, our limiting magnitudes for known objects from the last-night pipeline is $V$=22.9\,mag with the median value of $V$=20.4\,mag (almost 1\,mag fainter than WTS). One of the reasons of fewer detections in SPECULOOS data is longer exposure times which limit telescopes capabilities to discover fast moving objects. However, despite that and having a smaller aperture and FoV, SPECULOOS can compete with larger-aperture facilities such as UKIRT. Furthermore, taking into account results of the archival search for slow moving objects (which allowed detection of objects as faint as $V$=23.8\,mag), SPECULOOS can serve as pencil-beam surveys for small bodies and probe parameter space beyond $V$=22\,mag (see Fig.~\ref{fig:ap_mag_distr}), approaching limiting magnitudes from single \textit{HST} images \citep{2022A&A...661A..85K}.

\section{Conclusions}\label{sec:conclusions}

We have presented a data processing pipeline based on the publicly-available GPU-accelerated synthetic tracking software \textsc{Tycho Tracker}. We developed two versions of the pipeline. One version is designed to run daily after the end of an observing night and provides a list of all possible moving object candidates before the beginning of the next observing night (last-night data pipeline). Its development was motivated by our desire to be able to submit new observations to the MPC in a timely manner and to be able follow up objects of interest during the next observing night. The other version of the pipeline is intended to process archival images and is focused on detecting small bodies that move relatively slowly and could be present within the telescope's FoV for several subsequent nights (archival data pipeline). We analysed pipeline application on daily and archival data gathered by the SPECULOOS targeted ground-based photometric survey. 

\begin{figure}
    \centering
    \includegraphics[width=\columnwidth]{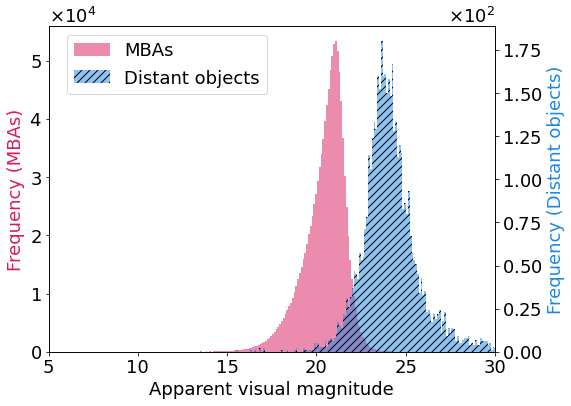}
    \caption{Apparent visual magnitude distribution of all Main-Belt Asteroids (MBAs) and distant objects (TNOs, Centaurs and Scattered Disk Objects) in the MPC database. Values are calculated for objects at opposition and using their absolute magnitudes $H$.}
    \label{fig:ap_mag_distr}
\end{figure}

Over the course of 60 nights, 90 different fields (3.6\,deg$^2$ in total) were observed with four SPECULOOS telescopes. In these last-night data sets, we identified 148 unique, known objects of different dynamical classes and 27 potentially new small bodies in 26 fields (1.04\,deg$^2$) located within 30\,deg from the ecliptic plane. From the search in the archival data which covers 390 individual nights, we identified 252 unique, known objects and 16 potentially new small bodies in 21 fields (0.84\,deg$^2$). All detections were made with no priors on the motion of small bodies and will be reported to the MPC. We were able to reach limiting magnitudes for unknown objects from the last-night and archival data of $V$=22.8\,mag and $V$=23.8\,mag, respectively. SPECULOOS and similar surveys can serve as pencil-beam surveys for small bodies and probe parameter space beyond $V$=22\,mag (see Fig.~\ref{fig:ap_mag_distr}). 

Three Near-Earth Asteroids (NEAs) were present in the last-night FoVs, but they were not detected because of their faintness ($V\geq23.7$\,mag). More objects were present in the archival data set (22 NEAs). None of them were detected as 20 objects were too faint (V > 24.1 mag) and remaining 2 brighter objects were too fast (speed ~1 arcsec/min) for archival configuration of the pipeline. We expect that relatively bright NEAs will be detected by the last-night version of the pipeline once more data is processed.

We were also able to recover small bodies using synthetic tracking in groups of images comprising several consecutive nights together. One of the detected objects (MBA 1999~BP4) was recovered only in 2-night sets because of its small apparent motion of only 0.01\,arcsec/min. This detection serves as a proof of concept for future application of our pipeline to detect TNOs and other distant objects
with a typical speed of 0.01\,arcsec/min and small daily changes of PA and angular speed.

For small bodies that persist within the FoV for an extended period of time, we have the opportunity to construct photometric light curves. Light curves can provide more detailed information about minor bodies, including rotation rates and surface characteristics. \textsc{Tycho Tracker} offers the capability to construct light curves and extract rotation periods across multiple nights of observations. In future work, we will use this capability to construct light curves of objects that remain in the fields over the course of more than one night. 

Our "blind" search is computationally intensive, but usage of even consumer-grade GPUs with a modest RAM of 8-10\,GB makes it possible to perform the search in reasonable time. 
A relatively quick processing time and ability to submit MPC reports by the last-night pipeline could allow for timely follow-up of NEAs and other types of small bodies and minimise losses of such objects. Additionally, the archival pipeline is able to produce numerous detections (tracklets) of small bodies, which will mostly populate the Isolated Tracklet File of the MPC. Such observations might be linked with previous or future observations done by other facilities and will contribute to a pool of objects with well-established orbits. And in some cases (when an object is present in an FoV for several days), its orbit can solely be established by the data from the archival pipeline. Over the course of a lifetime of the SPECULOOS project ($\sim$10 years), our pipelines will detect over ten thousand individual small bodies.

Ultimately, our data processing and analysis pipeline can be adapted and applied to other photometric surveys to search for moving bodies and maximize the overall science return. It might be especially promising in search for TNOs (which are predicted to outnumber any other type of small bodies) using JWST. Already, the pipeline has been adapted to analyse archival data from the ASTEP400 telescope, which is part of the ASTEP project (Antarctic Search for Transiting ExoPlanets; \citealt{daban2010ground}). Thanks to an uninterrupted night during the Antarctic winter and view of the ecliptic plane and solar antiapex \citep{hoover2022population}, this site is a good location to potentially discover interstellar objects (whose orbits are strongly hyperbolic and not gravitationally bound within the Solar System’s orbital plane) and Solar System minor bodies.

\section*{Acknowledgements}
\label{Ack}

J.d.W. and MIT gratefully acknowledge financial support from the Heising-Simons Foundation, Dr. and Mrs. Colin Masson and Dr. Peter A. Gilman for Artemis, the first telescope of the SPECULOOS network situated in Tenerife, Spain.

Authors would like to thank the reviewer for their time and attention. The constructive comments we received, helped us to improve the quality of the paper.

The authors would like to thank SPECULOOS consortium colleagues F.~Pozuelos, E.~Jehin, B.~Rackham, M.~Timmermans and P.~Niraula for their valuable comments. The authors would also like to thank developer of \textsc{Tycho Tracker} Daniel Parrott for his advice on the use of the software. 

This work has been supported by the NVIDIA Academic Hardware Grant Program. This research has made use of data and/or services provided by the International Astronomical Union's Minor Planet Center, the NASA/JPL Horizons On-Line Ephemeris System and Small-Body DataBase (\url{https://ssd.jpl.nasa.gov}). This research has made use of \textsc{NumPy} \citep{harris2020array}, \textsc{matplotlib} \citep{Hunter:2007}, \textsc{astropy} \citep{2013A&A...558A..33A, 2018AJ....156..123A}, \textsc{astroalign }\citep{beroiz2020astroalign}, \textsc{L.A.Cosmic} \citep{2001PASP..113.1420V} and \textsc{Tycho Tracker} software (\url{www.tycho-tracker.com}, \citealt{2020JAVSO..48..262P}).

\section*{Data availability}
\label{Data availability}

Data available on request.  The data underlying this article will be shared on reasonable request to the corresponding author.



\bibliographystyle{mnras}
\bibliography{luna_bibliography}

\begin{thebibliography}{}
\makeatletter
\relax
\def\mn@urlcharsother{\let\do\@makeother \do\$\do\&\do\#\do\^\do\_\do\%\do\~}
\def\mn@doi{\begingroup\mn@urlcharsother \@ifnextchar [ {\mn@doi@}
  {\mn@doi@[]}}
\def\mn@doi@[#1]#2{\def\@tempa{#1}\ifx\@tempa\@empty \href
  {http://dx.doi.org/#2} {doi:#2}\else \href {http://dx.doi.org/#2} {#1}\fi
  \endgroup}
\def\mn@eprint#1#2{\mn@eprint@#1:#2::\@nil}
\def\mn@eprint@arXiv#1{\href {http://arxiv.org/abs/#1} {{\tt arXiv:#1}}}
\def\mn@eprint@dblp#1{\href {http://dblp.uni-trier.de/rec/bibtex/#1.xml}
  {dblp:#1}}
\def\mn@eprint@#1:#2:#3:#4\@nil{\def\@tempa {#1}\def\@tempb {#2}\def\@tempc
  {#3}\ifx \@tempc \@empty \let \@tempc \@tempb \let \@tempb \@tempa \fi \ifx
  \@tempb \@empty \def\@tempb {arXiv}\fi \@ifundefined
  {mn@eprint@\@tempb}{\@tempb:\@tempc}{\expandafter \expandafter \csname
  mn@eprint@\@tempb\endcsname \expandafter{\@tempc}}}

\bibitem[\protect\citeauthoryear{{Astropy Collaboration} et~al.,}{{Astropy
  Collaboration} et~al.}{2013}]{2013A&A...558A..33A}
{Astropy Collaboration} et~al., 2013, \mn@doi [\aap]
  {10.1051/0004-6361/201322068}, \href
  {https://ui.adsabs.harvard.edu/abs/2013A&A...558A..33A} {558, A33}

\bibitem[\protect\citeauthoryear{{Astropy Collaboration} et~al.,}{{Astropy
  Collaboration} et~al.}{2018}]{2018AJ....156..123A}
{Astropy Collaboration} et~al., 2018, \mn@doi [\aj] {10.3847/1538-3881/aabc4f},
  \href {https://ui.adsabs.harvard.edu/abs/2018AJ....156..123A} {156, 123}

\bibitem[\protect\citeauthoryear{Bannister et~al.,}{Bannister
  et~al.}{2016}]{bannister2016outer}
Bannister M.~T.,  et~al., 2016, \aj, 152, 70

\bibitem[\protect\citeauthoryear{{Bernardinelli} et~al.,}{{Bernardinelli}
  et~al.}{2022}]{2022ApJS..258...41B}
{Bernardinelli} P.~H.,  et~al., 2022, \mn@doi [\apjs]
  {10.3847/1538-4365/ac3914}, \href
  {https://ui.adsabs.harvard.edu/abs/2022ApJS..258...41B} {258, 41}

\bibitem[\protect\citeauthoryear{Beroiz, Cabral  \& Sanchez}{Beroiz
  et~al.}{2020}]{beroiz2020astroalign}
Beroiz M.,  Cabral J.,   Sanchez B.,  2020, \mn@doi [Astron. Comput.]
  {https://doi.org/10.1016/j.ascom.2020.100384}, 32, 100384

\bibitem[\protect\citeauthoryear{{Bolin} et~al.,}{{Bolin}
  et~al.}{2022}]{2022MNRAS.517L..49B}
{Bolin} B.~T.,  et~al., 2022, \mn@doi [\mnras] {10.1093/mnrasl/slac089}, \href
  {https://ui.adsabs.harvard.edu/abs/2022MNRAS.517L..49B} {517, L49}

\bibitem[\protect\citeauthoryear{{Burdanov} et~al.,}{{Burdanov}
  et~al.}{2022}]{burdanov2022speculoos}
{Burdanov} A.~Y.,  et~al., 2022, \mn@doi [\pasp] {10.1088/1538-3873/ac92a6},
  \href {https://ui.adsabs.harvard.edu/abs/2022PASP..134j5001B} {134, 105001}

\bibitem[\protect\citeauthoryear{{Carry}}{{Carry}}{2018}]{2018A&A...609A.113C}
{Carry} B.,  2018, \mn@doi [\aap] {10.1051/0004-6361/201730386}, \href
  {https://ui.adsabs.harvard.edu/abs/2018A&A...609A.113C} {609, A113}

\bibitem[\protect\citeauthoryear{{Chen} et~al.,}{{Chen}
  et~al.}{2018}]{chen2018searching}
{Chen} Y.-T.,  et~al., 2018, \mn@doi [\pasj] {10.1093/pasj/psx145}, \href
  {https://ui.adsabs.harvard.edu/abs/2018PASJ...70S..38C} {70, S38}

\bibitem[\protect\citeauthoryear{Cochran, Levison, Stern  \& Duncan}{Cochran
  et~al.}{1995}]{cochran1995discovery}
Cochran A.~L.,  Levison H.~F.,  Stern S.~A.,   Duncan M.~J.,  1995, \mn@doi
  [\apj] {10.1086/176581}, 455, 342

\bibitem[\protect\citeauthoryear{{Cort{\'e}s-Contreras}
  et~al.,}{{Cort{\'e}s-Contreras} et~al.}{2019}]{2019MNRAS.490.3046C}
{Cort{\'e}s-Contreras} M.,  et~al., 2019, \mn@doi [\mnras]
  {10.1093/mnras/stz2727}, \href
  {https://ui.adsabs.harvard.edu/abs/2019MNRAS.490.3046C} {490, 3046}

\bibitem[\protect\citeauthoryear{{Daban} et~al.,}{{Daban}
  et~al.}{2010}]{daban2010ground}
{Daban} J.-B.,  et~al., 2010, in {Stepp} L.~M.,  {Gilmozzi} R.,   {Hall} H.~J.,
   eds,  Society of Photo-Optical Instrumentation Engineers (SPIE) Conference
  Series Vol. 7733, Ground-based and Airborne Telescopes III. p. 77334T,
  \mn@doi{10.1117/12.854946}

\bibitem[\protect\citeauthoryear{{DeMeo} \& {Carry}}{{DeMeo} \&
  {Carry}}{2014}]{2014Natur.505..629D}
{DeMeo} F.~E.,  {Carry} B.,  2014, \mn@doi [\nat] {10.1038/nature12908}, \href
  {https://ui.adsabs.harvard.edu/abs/2014Natur.505..629D} {505, 629}

\bibitem[\protect\citeauthoryear{{Delrez} et~al.,}{{Delrez}
  et~al.}{2018}]{2018SPIE10700E..1ID}
{Delrez} L.,  et~al., 2018, in {Marshall} H.~K.,  {Spyromilio} J.,  eds,
  Society of Photo-Optical Instrumentation Engineers (SPIE) Conference Series
  Vol. 10700, Ground-based and Airborne Telescopes VII. p. 107001I (\mn@eprint
  {arXiv} {1806.11205}), \mn@doi{10.1117/12.2312475}

\bibitem[\protect\citeauthoryear{{Delrez} et~al.,}{{Delrez}
  et~al.}{2022}]{2022A&A...667A..59D}
{Delrez} L.,  et~al., 2022, \mn@doi [\aap] {10.1051/0004-6361/202244041}, \href
  {https://ui.adsabs.harvard.edu/abs/2022A&A...667A..59D} {667, A59}

\bibitem[\protect\citeauthoryear{{Demory} et~al.,}{{Demory}
  et~al.}{2020}]{2020A&A...642A..49D}
{Demory} B.~O.,  et~al., 2020, \mn@doi [\aap] {10.1051/0004-6361/202038616},
  \href {https://ui.adsabs.harvard.edu/abs/2020A&A...642A..49D} {642, A49}

\bibitem[\protect\citeauthoryear{{Dones}, {Brasser}, {Kaib}  \&
  {Rickman}}{{Dones} et~al.}{2015}]{2015SSRv..197..191D}
{Dones} L.,  {Brasser} R.,  {Kaib} N.,   {Rickman} H.,  2015, \mn@doi [\ssr]
  {10.1007/s11214-015-0223-2}, \href
  {https://ui.adsabs.harvard.edu/abs/2015SSRv..197..191D} {197, 191}

\bibitem[\protect\citeauthoryear{Garcia, Timmermans, Pozuelos  et~al.}{Garcia
  et~al.}{2021}]{2022MNRAS.509.4817G}
Garcia L.~J.,  Timmermans M.,  Pozuelos F.~J.,   et~al., 2021, \mn@doi [\mnras]
  {10.1093/mnras/stab3113}, 509, 4817

\bibitem[\protect\citeauthoryear{{Gehrels} \& {McMillan}}{{Gehrels} \&
  {McMillan}}{1982}]{1982ASSL...96..279G}
{Gehrels} T.,  {McMillan} R.~S.,  1982, in {Fricke} W.,  {Teleki} G.,  eds,
  Astrophysics and Space Science Library Vol. 96, Sun and Planetary System.
  p.~279, \mn@doi{10.1007/978-94-009-7846-1_75}

\bibitem[\protect\citeauthoryear{{Gillon}}{{Gillon}}{2018}]{2018NatAs...2..344G}
{Gillon} M.,  2018, \mn@doi [Nat. Astron.] {10.1038/s41550-018-0443-y}, \href
  {https://ui.adsabs.harvard.edu/abs/2018NatAs...2..344G} {2, 344}

\bibitem[\protect\citeauthoryear{{Giorgini} et~al.,}{{Giorgini}
  et~al.}{1996}]{1996DPS....28.2504G}
{Giorgini} J.~D.,  et~al., 1996, in AAS/Division for Planetary Sciences Meeting
  Abstracts \#28. p. 25.04

\bibitem[\protect\citeauthoryear{{Gwyn}, {Hill}  \& {Kavelaars}}{{Gwyn}
  et~al.}{2012}]{2012PASP..124..579G}
{Gwyn} S. D.~J.,  {Hill} N.,   {Kavelaars} J.~J.,  2012, \mn@doi [\pasp]
  {10.1086/666462}, \href
  {https://ui.adsabs.harvard.edu/abs/2012PASP..124..579G} {124, 579}

\bibitem[\protect\citeauthoryear{Harris et~al.,}{Harris
  et~al.}{2020}]{harris2020array}
Harris C.~R.,  et~al., 2020, \mn@doi [\nat] {10.1038/s41586-020-2649-2}, 585,
  357

\bibitem[\protect\citeauthoryear{{Heinze}, {Metchev}  \& {Trollo}}{{Heinze}
  et~al.}{2015}]{2015AJ....150..125H}
{Heinze} A.~N.,  {Metchev} S.,   {Trollo} J.,  2015, \mn@doi [\aj]
  {10.1088/0004-6256/150/4/125}, \href
  {https://ui.adsabs.harvard.edu/abs/2015AJ....150..125H} {150, 125}

\bibitem[\protect\citeauthoryear{{H{\o}g} et~al.,}{{H{\o}g}
  et~al.}{2000}]{2000A&A...355L..27H}
{H{\o}g} E.,  et~al., 2000, \aap, \href
  {https://ui.adsabs.harvard.edu/abs/2000A&A...355L..27H} {355, L27}

\bibitem[\protect\citeauthoryear{{Holler} et~al.,}{{Holler}
  et~al.}{2018}]{2018JATIS...4c4003H}
{Holler} B.~J.,  et~al., 2018, \mn@doi [Journal of Astronomical Telescopes,
  Instruments, and Systems] {10.1117/1.JATIS.4.3.034003}, \href
  {https://ui.adsabs.harvard.edu/abs/2018JATIS...4c4003H} {4, 034003}

\bibitem[\protect\citeauthoryear{{Holman}, {Payne}, {Blankley}, {Janssen}  \&
  {Kuindersma}}{{Holman} et~al.}{2018}]{2018arXiv180502638H}
{Holman} M.~J.,  {Payne} M.~J.,  {Blankley} P.,  {Janssen} R.,   {Kuindersma}
  S.,  2018, arXiv e-prints, \href
  {https://ui.adsabs.harvard.edu/abs/2018arXiv180502638H} {p. arXiv:1805.02638}

\bibitem[\protect\citeauthoryear{Hoover, Seligman  \& Payne}{Hoover
  et~al.}{2022}]{hoover2022population}
Hoover D.~J.,  Seligman D.~Z.,   Payne M.~J.,  2022, The Planetary Science
  Journal, 3, 71

\bibitem[\protect\citeauthoryear{Hunter}{Hunter}{2007}]{Hunter:2007}
Hunter J.~D.,  2007, \mn@doi [Computing in Science \& Engineering]
  {10.1109/MCSE.2007.55}, 9, 90

\bibitem[\protect\citeauthoryear{{Ivezi{\'c}} et~al.,}{{Ivezi{\'c}}
  et~al.}{2001}]{2001AJ....122.2749I}
{Ivezi{\'c}} {\v{Z}}.,  et~al., 2001, \mn@doi [\aj] {10.1086/323452}, \href
  {https://ui.adsabs.harvard.edu/abs/2001AJ....122.2749I} {122, 2749}

\bibitem[\protect\citeauthoryear{{Jehin} et~al.,}{{Jehin}
  et~al.}{2018}]{2018Msngr.174....2J}
{Jehin} E.,  et~al., 2018, \mn@doi [The Messenger] {10.18727/0722-6691/5105},
  \href {https://ui.adsabs.harvard.edu/abs/2018Msngr.174....2J} {174, 2}

\bibitem[\protect\citeauthoryear{Kaiser et~al.,}{Kaiser
  et~al.}{2002}]{kaiser2002pan}
Kaiser N.,  et~al., 2002, in Survey and Other Telescope Technologies and
  Discoveries. pp 154--164

\bibitem[\protect\citeauthoryear{{Kruk} et~al.,}{{Kruk}
  et~al.}{2022}]{2022A&A...661A..85K}
{Kruk} S.,  et~al., 2022, \mn@doi [\aap] {10.1051/0004-6361/202142998}, \href
  {https://ui.adsabs.harvard.edu/abs/2022A&A...661A..85K} {661, A85}

\bibitem[\protect\citeauthoryear{{LSST Science Collaboration} et~al.,}{{LSST
  Science Collaboration} et~al.}{2009}]{2009arXiv0912.0201L}
{LSST Science Collaboration} et~al., 2009, arXiv e-prints, \href
  {https://ui.adsabs.harvard.edu/abs/2009arXiv0912.0201L} {p. arXiv:0912.0201}

\bibitem[\protect\citeauthoryear{{Lang}, {Hogg}, {Mierle}, {Blanton}  \&
  {Roweis}}{{Lang} et~al.}{2010}]{2010AJ....139.1782L}
{Lang} D.,  {Hogg} D.~W.,  {Mierle} K.,  {Blanton} M.,   {Roweis} S.,  2010,
  \mn@doi [\aj] {10.1088/0004-6256/139/5/1782}, \href
  {https://ui.adsabs.harvard.edu/abs/2010AJ....139.1782L} {139, 1782}

\bibitem[\protect\citeauthoryear{Larson, Brownlee, Hergenrother  \&
  Spahr}{Larson et~al.}{1998}]{larson1998catalina}
Larson S.,  Brownlee J.,  Hergenrother C.,   Spahr T.,  1998, in \baas. p.~1037

\bibitem[\protect\citeauthoryear{{Mainzer} \& {NEOCam Science Team}}{{Mainzer}
  \& {NEOCam Science Team}}{2017}]{2017DPS....4921901M}
{Mainzer} A.~K.,  {NEOCam Science Team} 2017, in AAS/Division for Planetary
  Sciences Meeting Abstracts \#49. p. 219.01

\bibitem[\protect\citeauthoryear{{Marton} et~al.,}{{Marton}
  et~al.}{2020}]{2020Icar..34513721M}
{Marton} G.,  et~al., 2020, \mn@doi [\icarus] {10.1016/j.icarus.2020.113721},
  \href {https://ui.adsabs.harvard.edu/abs/2020Icar..34513721M} {345, 113721}

\bibitem[\protect\citeauthoryear{{McCully} et~al.,}{{McCully}
  et~al.}{2018}]{curtis_mccully_2018_1482019}
{McCully} C.,  et~al., 2018, astropy/astroscrappy: v1.0.5 Zenodo Release,
  \mn@doi{10.5281/zenodo.1482019}, \url
  {https://doi.org/10.5281/zenodo.1482019}

\bibitem[\protect\citeauthoryear{Michel, DeMeo  \& Bottke}{Michel
  et~al.}{2015}]{michel2015asteroids}
Michel P.,  DeMeo F.~E.,   Bottke W.~F.,  2015, Asteroids IV, 1, 1

\bibitem[\protect\citeauthoryear{{Murray} et~al.,}{{Murray}
  et~al.}{2020}]{2020MNRAS.495.2446M}
{Murray} C.~A.,  et~al., 2020, \mn@doi [\mnras] {10.1093/mnras/staa1283}, \href
  {https://ui.adsabs.harvard.edu/abs/2020MNRAS.495.2446M} {495, 2446}

\bibitem[\protect\citeauthoryear{{Niraula} et~al.,}{{Niraula}
  et~al.}{2020}]{Niraula2020}
{Niraula} P.,  et~al., 2020, \mn@doi [\aj] {10.3847/1538-3881/aba95f}, \href
  {https://ui.adsabs.harvard.edu/abs/2020AJ....160..172N} {160, 172}

\bibitem[\protect\citeauthoryear{Osinski, Cockell, Pontefract  \&
  Sapers}{Osinski et~al.}{2020}]{osinski2020role}
Osinski G.,  Cockell C.,  Pontefract A.,   Sapers H.,  2020, Astrobiology, 20,
  1121

\bibitem[\protect\citeauthoryear{Parker \& Kavelaars}{Parker \&
  Kavelaars}{2010}]{parker2010pencil}
Parker A.~H.,  Kavelaars J.,  2010, \pasp, 122, 549

\bibitem[\protect\citeauthoryear{{Parrott}}{{Parrott}}{2020}]{2020JAVSO..48..262P}
{Parrott} D.,  2020, The Journal of the American Association of Variable Star
  Observers, \href {https://ui.adsabs.harvard.edu/abs/2020JAVSO..48..262P} {48,
  262}

\bibitem[\protect\citeauthoryear{{Popescu} et~al.,}{{Popescu}
  et~al.}{2016}]{2016A&A...591A.115P}
{Popescu} M.,  et~al., 2016, \mn@doi [\aap] {10.1051/0004-6361/201628163},
  \href {https://ui.adsabs.harvard.edu/abs/2016A&A...591A.115P} {591, A115}

\bibitem[\protect\citeauthoryear{{Pravdo} et~al.,}{{Pravdo}
  et~al.}{1999}]{1999AJ....117.1616P}
{Pravdo} S.~H.,  et~al., 1999, \mn@doi [\aj] {10.1086/300769}, \href
  {https://ui.adsabs.harvard.edu/abs/1999AJ....117.1616P} {117, 1616}

\bibitem[\protect\citeauthoryear{{Rice} \& {Laughlin}}{{Rice} \&
  {Laughlin}}{2020}]{2020PSJ.....1...81R}
{Rice} M.,  {Laughlin} G.,  2020, \mn@doi [Planetary Science Journal]
  {10.3847/PSJ/abc42c}, \href
  {https://ui.adsabs.harvard.edu/abs/2020PSJ.....1...81R} {1, 81}

\bibitem[\protect\citeauthoryear{{Ricker} et~al.,}{{Ricker}
  et~al.}{2015}]{2015JATIS...1a4003R}
{Ricker} G.~R.,  et~al., 2015, \mn@doi [Journal of Astronomical Telescopes,
  Instruments, and Systems] {10.1117/1.JATIS.1.1.014003}, \href
  {https://ui.adsabs.harvard.edu/abs/2015JATIS...1a4003R} {1, 014003}

\bibitem[\protect\citeauthoryear{{Schanche} et~al.,}{{Schanche}
  et~al.}{2022}]{2022A&A...657A..45S}
{Schanche} N.,  et~al., 2022, \mn@doi [\aap] {10.1051/0004-6361/202142280},
  \href {https://ui.adsabs.harvard.edu/abs/2022A&A...657A..45S} {657, A45}

\bibitem[\protect\citeauthoryear{{Sebastian} et~al.,}{{Sebastian}
  et~al.}{2021}]{2021A&A...645A.100S}
{Sebastian} D.,  et~al., 2021, \mn@doi [\aap] {10.1051/0004-6361/202038827},
  \href {https://ui.adsabs.harvard.edu/abs/2021A&A...645A.100S} {645, A100}

\bibitem[\protect\citeauthoryear{{Sergeyev} \& {Carry}}{{Sergeyev} \&
  {Carry}}{2021}]{2021A&A...652A..59S}
{Sergeyev} A.~V.,  {Carry} B.,  2021, \mn@doi [\aap]
  {10.1051/0004-6361/202140430}, \href
  {https://ui.adsabs.harvard.edu/abs/2021A&A...652A..59S} {652, A59}

\bibitem[\protect\citeauthoryear{Shao, Nemati, Zhai, Turyshev, Sandhu, Hallinan
   \& Harding}{Shao et~al.}{2014}]{shao2014finding}
Shao M.,  Nemati B.,  Zhai C.,  Turyshev S.~G.,  Sandhu J.,  Hallinan G.,
  Harding L.~K.,  2014, \apj, 782, 1

\bibitem[\protect\citeauthoryear{Sheppard, Trujillo  \& Tholen}{Sheppard
  et~al.}{2019}]{sheppard2019probing}
Sheppard S.,  Trujillo C.,   Tholen D.,  2019, in EPSC-DPS Joint Meeting 2019.
  pp EPSC--DPS2019

\bibitem[\protect\citeauthoryear{{Space Studies Board}, {National Research
  Council}  et~al.}{{Space Studies Board} et~al.}{2010}]{board2010defending}
{Space Studies Board} {National Research Council}  et~al., 2010, Defending
  planet earth: Near-Earth-Object surveys and hazard mitigation strategies.
National Academies Press

\bibitem[\protect\citeauthoryear{Stokes, Evans, Viggh, Shelly  \&
  Pearce}{Stokes et~al.}{2000}]{stokes2000lincoln}
Stokes G.~H.,  Evans J.~B.,  Viggh H.~E.,  Shelly F.~C.,   Pearce E.~C.,  2000,
  \icarus, 148, 21

\bibitem[\protect\citeauthoryear{{Tanga} et~al.,}{{Tanga}
  et~al.}{2022}]{2022arXiv220605561T}
{Tanga} P.,  et~al., 2022, arXiv e-prints, \href
  {https://ui.adsabs.harvard.edu/abs/2022arXiv220605561T} {p. arXiv:2206.05561}

\bibitem[\protect\citeauthoryear{Tyson, Guhathakurta, Bernstein  \& Hut}{Tyson
  et~al.}{1992}]{tyson1992limits}
Tyson J.,  Guhathakurta P.,  Bernstein G.,   Hut P.,  1992, in American
  Astronomical Society Meeting Abstracts. pp 06--10

\bibitem[\protect\citeauthoryear{Vaduvescu, Curelaru, Birlan, Bocsa,
  Serbanescu, Tudorica  \& Berthier}{Vaduvescu
  et~al.}{2009}]{vaduvescu2009euronear}
Vaduvescu O.,  Curelaru L.,  Birlan M.,  Bocsa G.,  Serbanescu L.,  Tudorica
  A.,   Berthier J.,  2009, Astronomische Nachrichten: Astronomical Notes, 330,
  698

\bibitem[\protect\citeauthoryear{{Vaduvescu}, {Curelaru}  \&
  {Popescu}}{{Vaduvescu} et~al.}{2020}]{2020A&C....3000356V}
{Vaduvescu} O.,  {Curelaru} L.,   {Popescu} M.,  2020, \mn@doi [Astronomy and
  Computing] {10.1016/j.ascom.2019.100356}, \href
  {https://ui.adsabs.harvard.edu/abs/2020A&C....3000356V} {30, 100356}

\bibitem[\protect\citeauthoryear{{Vere{\v{s}}}, {Payne}, {Holman},
  {Farnocchia}, {Williams}, {Keys}  \& {Boardman}}{{Vere{\v{s}}}
  et~al.}{2018}]{2018AJ....156....5V}
{Vere{\v{s}}} P.,  {Payne} M.~J.,  {Holman} M.~J.,  {Farnocchia} D.,
  {Williams} G.~V.,  {Keys} S.,   {Boardman} I.,  2018, \mn@doi [\aj]
  {10.3847/1538-3881/aac37d}, \href
  {https://ui.adsabs.harvard.edu/abs/2018AJ....156....5V} {156, 5}

\bibitem[\protect\citeauthoryear{{Wells} et~al.,}{{Wells}
  et~al.}{2021}]{Wells2021}
{Wells} R.~D.,  et~al., 2021, \mn@doi [\aap] {10.1051/0004-6361/202141277},
  \href {https://ui.adsabs.harvard.edu/abs/2021A&A...653A..97W} {653, A97}

\bibitem[\protect\citeauthoryear{Weryk et~al.,}{Weryk
  et~al.}{2016}]{weryk2016distant}
Weryk R.,  et~al., 2016, arXiv preprint arXiv:1607.04895

\bibitem[\protect\citeauthoryear{{Woods} et~al.,}{{Woods}
  et~al.}{2021}]{2021PASP..133a4503W}
{Woods} D.~F.,  et~al., 2021, \mn@doi [\pasp] {10.1088/1538-3873/abc761}, \href
  {https://ui.adsabs.harvard.edu/abs/2021PASP..133a4503W} {133, 014503}

\bibitem[\protect\citeauthoryear{Wright et~al.,}{Wright
  et~al.}{2010}]{wright2010wide}
Wright E.~L.,  et~al., 2010, \aj, 140, 1868

\bibitem[\protect\citeauthoryear{{Zhai} et~al.,}{{Zhai}
  et~al.}{2020}]{2020PASP..132f4502Z}
{Zhai} C.,  et~al., 2020, \mn@doi [\pasp] {10.1088/1538-3873/ab828b}, \href
  {https://ui.adsabs.harvard.edu/abs/2020PASP..132f4502Z} {132, 064502}

\bibitem[\protect\citeauthoryear{{van Dokkum}}{{van
  Dokkum}}{2001}]{2001PASP..113.1420V}
{van Dokkum} P.~G.,  2001, \mn@doi [\pasp] {10.1086/323894}, \href
  {https://ui.adsabs.harvard.edu/abs/2001PASP..113.1420V} {113, 1420}

\makeatother
\end{thebibliography}


\phantom{invisible text}\\

\begin{table*}
\begin{center}

    \caption{Fields with small body detections observed by the four SPECULOOS telescopes from mid-September to mid-November 2022 and processed by the last-night version of the pipeline. Latitude and longitude coordinates are provided in the geocentric mean ecliptic reference frame. Faintest $\mathrm{V_{mag}}$ refers to a visual magnitude of the faintest detected known object in the field (if no known objects were detected, $\mathrm{V_{mag}}$ refers to the unknown object).} \label{last-night_search_fields}

\begin{tabular}{lccccccccc}
\hline
Field name  & \begin{tabular}[c]{@{}c@{}}Longitude \\(deg)\end{tabular} & \begin{tabular}[c]{@{}c@{}}Latitude \\(deg)\end{tabular} & \begin{tabular}[c]{@{}c@{}}RA \\(hms)\end{tabular} & \begin{tabular}[c]{@{}c@{}}Dec \\(dms)\end{tabular} & Filter & \begin{tabular}[c]{@{}c@{}}Observed \\ hours\\ \end{tabular} & \begin{tabular}[c]{@{}c@{}}Known \\ objects \end{tabular} & \begin{tabular}[c]{@{}c@{}}Unknown \\objects\end{tabular} & \begin{tabular}[c]{@{}c@{}}Faintest \\ $\mathrm{V_{mag}}$ \end{tabular} \\ \hline
Sp0004-2058 & 352.379 & -19.635 & 00 04 42 & -20 58 30 & $I+z'$ & 3 & - & 1 & 21.8 \\
Sp0006-0732 & 358.521 & -7.581 & 00 06 43 & -7 32 17 & $r'$ & 13 & 9 & 3 & 20.9 \\
Sp0109-0343 & 14.651 & -10.295 & 01 09 51 & -3 43 26 & $I+z'$ & 15 & 16 & 1 & 21.5\\
Sp0202+1020 & 32.049 & -1.97 & 02 02 16 & 10 20 14 & $i'$ & 6 & 4 & - & 19.8\\
Sp0216+1335 & 36.419 & -0.079 & 02 16 30 & 13 35 13 & $i'$ & 5 & 11 & - & 21.3\\
Sp0248-1651 & 33.757 & -31.441 & 02 48 41 & -16 51 22 & $I+z'$ & 17 & 2 & - & 22.3\\
Sp0253+1652 & 45.814 & 0.321 & 02 53 01 & 16 52 53 & $r'$ & 15 & 12 & 2 & 22.1\\
Sp0314+1603 & 50.432 & -1.873 & 03 14 03 & 16 03 06 & $I+z'$ & 7 & 10 & 4 & 21.7 \\
Sp0330+5413 & 64.699 & 34.02 & 03 30 49 & 54 13 55 & $i'$ & 21 & 2 & 1 & 20.4\\
Sp0351-0052 & 55.268 & -20.515 & 03 51 00 & 00 52 45 & $I+z'$ & 13 & 2 & 1 & 18.0 \\
Sp0409-0605 & 58.919 & -26.593 & 04 09 22 & -6 05 19 & $I+z'$ & 18 & 1 & 1 & 18.5\\
Sp0426+0336 & 65.37 & -17.85 & 04 26 20 & 03 36 36 & $I+z'$ & 11 & 1 & - & 22.7\\
Sp0510+2714 & 78.943 & 4.27 & 05 10 20 & 27 14 02 & $z'$ & 8 & 10 & - & 20.3\\
Sp0602+3910 & 90.501 & 15.745 & 06 02 31 & 39 10 59 & $I+z'$ & 40 & 4 & - & 20.1 \\
Sp0602-0915 & 90.848 & -32.688 & 06 02 54 & -9 15 04 & $I+z'$ & 8 & 1 & - & 20.5\\
Sp0714+3702 & 105.179 & 14.578 & 07 14 04 & 37 02 46 & $I+z'$ & 31 & 3 & 1 & 21.5\\
Sp2327-1741 & 345.454 & -13.013 & 23 27 26 & -17 41 33 & $I+z'$ & 28 & 5 & 2 & 20.7\\
Sp2335-0223 & 353.353 & 0.271 & 23 35 10 & -2 23 21 & $r'$ & 7 & 5 & 1 & 20.6\\
Sp2346+1129 & 1.596 & 11.841 & 23 46 46 & 11 29 09 & $I+z'$ & 47 & 8 & 5 & 21.3\\
Sp2354-3316 & 344.114 & -29.662 & 23 54 09 & -33 16 27 & $I+z'$ & 25 & - & 1 & 21.4 \\
TOI-3714 & 73.597 & 17.176 & 04 38 13 & 39 27 29 & $g'$ & 7 & 1 & - & 21.1 \\
TOI-4506 & 349.742 & -9.564 & 23 37 39 & -12 50 33 & $g'$ & 1 & 1 & - & 21.4\\
TOI-5315 & 28.346 & 5.5 & 01 37 12 & 16 01 07 & $r'$ & 11 & 2 & - & 21.5\\
TOI-5349 & 55.503 & 1.794 & 03 30 51 & 20 52 47 & $r'$ & 10 & 4 & 3 & 21.5\\
HIP~41378 & 122.006 & -9.87 & 08 08 08 & 10 04 49 & $r'$ & 1 & 1 & - & 21.1\\
TRAPPIST-1 & 345.729 & 0.634 & 23 06 29 & -5 02 28 & $I+z'$ & 46 & 33 & - & 22.1\\                                            
\end{tabular}

\end{center}

\end{table*}
\begin{table*}
\begin{center}

    \caption{All sky fields processed by the archival version of the pipeline. All fields are located within 5\,deg from the ecliptic plane and were observed for more than 5 hours by five SPECULOOS telescopes from 2018 until mid-2022. Latitude and longitude coordinates are provided in the geocentric mean ecliptic reference frame. Faintest $\mathrm{V_{mag}}$ refers to a visual magnitude of the faintest detected known object in the field (if no known objects were detected, $\mathrm{V_{mag}}$ refers to the unknown object).}\label{slow_search_fields}

\begin{tabular}{lccccccccc}
\hline
Field name  & \begin{tabular}[c]{@{}c@{}}Longitude \\(deg)\end{tabular} & \begin{tabular}[c]{@{}c@{}}Latitude \\(deg)\end{tabular} & \begin{tabular}[c]{@{}c@{}}RA \\(hms)\end{tabular} & \begin{tabular}[c]{@{}c@{}}Dec \\(dms)\end{tabular} & Filter & \begin{tabular}[c]{@{}c@{}}Observed \\ hours\\ \end{tabular} & \begin{tabular}[c]{@{}c@{}}Known \\ objects \end{tabular} & \begin{tabular}[c]{@{}c@{}}Unknown \\objects\end{tabular}  & \begin{tabular}[c]{@{}c@{}}Faintest \\ $\mathrm{V_{mag}}$ \end{tabular} \\ \hline
Sp0024-0158 &   4.819     &     --4.235     & 00 24 24      & --01 58 19   & $I+z'$  & 81  &  20 & 6 & 23.5  \\
Sp0202+1020 &   32.055     &    --1.970      & 02 02 16      & 10 20 13     & $i'$    & 37  & 9  &  - & 22.8 \\
Sp0216+1335 &    36.426   &   --0.079       & 02 16 29      & 13 35 12     & $i'$     & 35  &  1 &  - & 22.0  \\
Sp0253+1652 &   45.821    &     0.321       & 02 53 01      & 16 52 52     & $I+z'$     & 10  & 17 &  - & 23.0 \\
Sp0314+1603 &   50.439    &     --1.873     & 03 14 03      & 16 03 05     &   $I+z'$    & 56  &  7 & 4 & 22.8 \\
Sp0320+1854 &   52.777     &     0.456      & 03 20 59      & 18 54 22     & $I+z'$     &  39 &  6 &  - & 23.0 \\
Sp0510+2714 &   78.953      &     4.270      & 05 10 20      & 27 14 01     &  $I+z'$    & 60  & 30  &  - & 23.4  \\
Sp0741+1738 &   114.071     &     --3.706     & 07 41 06      & 17 38 44     &  $I+z'$    &  22 & -  &  - & - \\
Sp0752+1612 &   116.992     &     --4.641     & 07 52 23    & 16 12 15     &  $I+z'$    &  10 & 3  &  - & 22.3 \\
Sp0840+1824 &   127.702     &     0.059       & 08 40 29    & 18 24 08     & $I+z'$    & 53  &  17 & 3 & 23.0  \\
Sp0900+2150 &   131.278     &     4.622       & 09 00 23    & 21 50 04     &   $i'$   & 19 &  2 &  - & 22.9 \\
Sp0949+0806 &   146.763     &     --4.762     & 09 49 22    & 08 06 45     &  $I+z'$   &  82 & 8  &  - & 23.1 \\
Sp1056+0700 &   162.681     &     0.230       & 10 56 28   & 07 00 51      &   $r'$   & 9  & 3  &  - & 20.5 \\
Sp1444-2019 &   224.866     &     --4.215     & 14 44 20   & --20 19 22    &  $I+z'$   &  55 & 15 &  2 & 23.4 \\
Sp1507-2000 &   229.982     &     --2.363      & 15 07 27   & --20 00 43    &  $I+z'$    &  62 &  33 & - & 23.5 \\
Sp1534-1418 &   234.885     &     4.815       & 15 34 56   & --14 18 49    &  $I+z'$   & 110  & 18  & 1 & 23.3  \\
Sp1656-2046 &   255.183      & 1.850           & 16 56 33   & --20 46 37    &  $I+z'$   & 66  &  13 &  - & 22.7  \\
Sp1845-2535 &   280.168     & --2.545         & 18 45 07   & --25 35 13    &  $I+z'$   & 94  &  4 & - & 21.2 \\
Sp2049-1716 &   310.144     & 0.448            & 20 49 52   & --17 16 08    &  $I+z'$   & 173  & 23  &  - & 23.2 \\
Sp2210-0936 &   331.249     & 1.528           & 22 10 55   & --09 36 01    &  $I+z'$    & 38  &  6 &  - & 22.6 \\
Sp2228-1325 &   333.978     & --3.624         & 22 28 54   & --13 25 19    &  $I+z'$   &  19 & 13  &  - & 23.0 \\
Sp2315-0627 &   347.338     & --1.587         & 23 15 54   & --06 27 46    &  $I+z'$    &  49 & 4  &  - & 22.5                                                        
\end{tabular}

\end{center}

\end{table*}


\bsp	
\label{lastpage}
\end{document}